\documentclass[aps,prl,reprint,groupedaddress,superscriptaddress,amsfonts,amssymb,amsmath,showpacs,floatfix]{revtex4-1}
\usepackage{color}
\usepackage{graphicx}

\newcommand{\dxspiral}{\texttt{dxspiral}} 
\newcommand{\ezride}{\texttt{EZRide}} 
\newcommand{\eg}[1]{{e.g.}\ifx#1.\else\expandafter#1\fi}
\newcommand{\ie}[1]{{i.e.}\ifx#1.\else\expandafter#1\fi}

\newcommand{\Complex}{\mathbb{C}}       
\newcommand{\const}{\mathrm{const}}     
\renewcommand{\d}{\mathrm{d}}           
\newcommand{\dd}{\partial}              
\newcommand{\df}[2]{\frac{\partial #1}{\partial #2}} 
\newcommand{\diag}{\mathrm{diag}}       
\newcommand{\adj}{^{\dagger}}           
\renewcommand{\H}{^{H}}                 
\renewcommand{\Im}[1]{\mathop{\mathrm{Im}}\left(#1\right)} 
\newcommand{\Integer}{\mathbb{Z}}       
\newcommand{\mx}[1]{\mathbf{#1}}        
\newcommand{\OO} {\mathcal {O}}         
\renewcommand{\Re}[1]{\mathop{\mathrm{Re}}\left(#1\right)} 
\newcommand{\Real}{\mathbb{R}}          
\newcommand{\T}{^{\mathrm{T}}}          

\newcommand{\bra}[1]{\mathinner{\langle{#1}|}}              
\newcommand{\ket}[1]{\mathinner{|{#1}\rangle}}              
\newcommand{\braket}[2]{\mathinner{\langle{#1|#2}\rangle}}  

\renewcommand{\AA}{\hat{\mathbf{A}}}    
\newcommand{\Amp}{A}             
\newcommand{\Ax}{A_x}            
\newcommand{\Ay}{A_y}            
\newcommand{\Axy}{A_{x,y}}       
\newcommand{\avg}[1]{\left\langle#1\right\rangle} 
\newcommand{\avec}{\mathbf{a}}   
\newcommand{\bvec}{\mathbf{b}}   
\newcommand{\apar}{a}            
\newcommand{\bpar}{b}            
\newcommand{\cpar}{c}            
\newcommand{\Dv}{D_v}            
\newcommand{\D}{\mx{D}}          
\newcommand{\f}{\mx{f}}          
\newcommand{\dt}{\Delta_t}       
\newcommand{\dx}{\Delta_x}       
\newcommand{\HL}{\hat{\mathbf{L}}} 
\newcommand{\HP}{\D}             
\newcommand{\hpi}{\hat{\pi}}     
\newcommand{\kz}{k_z}            
\newcommand{\kc}{k_*}            
\newcommand{\ko}{k_0}            
\newcommand{\Lx}{L_x}            
\newcommand{\Ly}{L_y}            
\newcommand{\Lz}{L_z}            
\newcommand{\Lc}{L_*}            
\newcommand{\Li}{\HL'}           
\newcommand{\la}{\lambda}        
\newcommand{\lamv}{\lambda}      
\newcommand{\Nr}{N_{\rho}}       
\newcommand{\Nt}{N_{\theta}}     
\newcommand{\Nz}{N_z}            
\newcommand{\nonsc}{b_1}         
\newcommand{\nonps}{b_2}         
\newcommand{\Om}{\Omega}         
\newcommand{\Omc}{\Omega_*}      
\newcommand{\om}{\omega}         
\newcommand{\Phase}{\Phi}        
\newcommand{\R}{\vec{R}}         
\newcommand{\rp}{\rho_{\max}}    
\renewcommand{\r}{\vec{r}}       
\newcommand{\rig}{e}             
\newcommand{\rigsc}{e_1}         
\newcommand{\rigps}{e_2}         
\newcommand{\s}{\sigma}          
\newcommand{\ten}{\gamma}        
\newcommand{\tensc}{\gamma_1}    
\newcommand{\tenps}{\gamma_2}    
\newcommand{\tuu}{\tilde{\u}}    
\newcommand{\UU}{\mathbf{U}}     
\renewcommand{\u}{\mx{u}}        
\newcommand{\uc}{u_*}            
\newcommand{\Vspace}{\mathcal{V}}
\newcommand{\VV}{\mathbf{V}}     
\newcommand{\vv}{\mathbf{v}}     
\newcommand{\vc}{v_*}            
\newcommand{\Wspace}{\mathcal{W}}
\newcommand{\WW}{\mathbf{W}}     
\newcommand{\ww}{\mathbf{w}}     
\newcommand{\xr}{x_r}            
\newcommand{\yr}{y_r}            
\newcommand{\zr}{z_r}            

\newcommand{\Fig}[1]{Fig.~\ref{fig:#1}}
\newcommand{\Figs}[1]{Figures~\ref{fig:#1}}
\newcommand{\fig}[1]{fig.~\ref{fig:#1}}
\newcommand{\figs}[1]{figures~\ref{fig:#1}}
\newcommand{\figref}[1]{\ref{fig:#1}}
\newcommand{\sglfigure}[3]{ 
  \begin{figure}[tb!]
  \centerline{\includegraphics{#1}}
  \caption[]{#2}
  \label{fig:#3}
  \end{figure}
}
\newcommand{\dblfigure}[3]{ 
  \begin{figure*}[tb!]
  \centerline{\includegraphics{#1}}
  \caption[]{#2}
  \label{fig:#3}
  \end{figure*}
}

\newcommand{\eq}[1]{\eqref{#1}}         
\newcommand{\nn}{\nonumber}             

\begin{document}
\title{Buckling of scroll waves}
\author{Hans Dierckx}
\author{Henri Verschelde}
\affiliation{Department of Mathematical Physics and Astronomy, 
  Ghent University, Krijgslaan 281, 9000 Ghent, Belgium}
\author{\"Ozg\"ur Selsil}
\affiliation{Department of Mathematical Sciences, University of Liverpool, Liverpool L69 7ZL, UK}
\author{Vadim N. Biktashev}
\affiliation{College of Engineering, Mathematics and Physical Sciences, University of Exeter, Exeter EX4 4QF, UK}
\date{\today}

\begin{abstract}
  A scroll wave in a sufficiently thin layer of an excitable medium
  with negative filament tension can be stable nevertheless
  due to filament rigidity. Above a certain critical thickness of the
  medium, such scroll wave will have a tendency to deform into a buckled,
  precessing state. Experimentally this will be seen as meandering of
  the spiral wave on the surface, the amplitude of which grows with
  the thickness of the layer, until a break-up to scroll wave
  turbulence happens. We present a simplified theory for this
  phenomenon and illustrate it with numerical examples.
\end{abstract}

\maketitle

Spiral waves in two-dimensions, and scroll waves in three-dimensions,
are regimes of self-organization observed in physical, chemical and
biological dissipative systems, where wave propagation is supported by
a source of energy stored in the medium \cite{%
  Zhabotinsky-Zaikin-1971,
  *Allessie-etal-1973,
  *Alcantara-Monk-1974,
  *Gorelova-Bures-1983,
  *Madore-Freedman-1987,
  *Jakubith-etal-1990,
  *Lechleiter-etal-1991,
  *Frisch-etal-1994,
  Cross-Hohenberg-1993%
}.
Due to effective localization of the critical adjoint eigenfunctions,
or ``response
functions''~\cite{Biktashev-etal-1994,Biktasheva-etal-1998}, the
dynamics of a spiral wave can be asymptotically described as that of
pointwise objects, in terms of its instant rotation centre and
phase~\cite{Biktashev-Holden-1995}.  The third dimension endows
scrolls with extra degrees of freedom: the filaments, around which the
scroll waves rotate, can bend, and the phase of rotation may vary
along the filaments, giving scrolls a
twist~\cite{Winfree-Strogatz-1983}.  The localization of response
functions allows description of scroll waves as string-like
objects~\cite{%
  Yakushevich-1984,%
  Brazhnik-etal-1987,%
  Keener-1988,%
  Biktashev-etal-1994,%
  Verschelde-2007,%
  Dierckx-2010%
}.  One manifestation of the extra degrees of freedom is the
possibility of ``scroll wave turbulence'' due to negative tension of
filaments~\cite{Biktashev-1998}. It has been speculated that this
scroll wave turbulence is in some respects similar to the hydrodynamic
turbulence, and may provide insights into the mechanisms of cardiac
fibrillation~\cite{Biktashev-etal-1994,Winfree-1994,Biktashev-1998,Zaritsky-etal-2004}.

The motivation for the present study comes from the analogy with
hydrodynamics.  At intermediate Reynolds
numbers, laminar flow can be unstable, leading to non-stationary
regimes which  are not 
turbulent~\cite{Cross-Hohenberg-1993}.  The possibility of similar
pre-turbulent regimes in scroll waves is interesting, \eg\ in view of
its possible relevance to cardiac arrhythmias. Cardiac muscle may be
considered quasi-two-dimensional if it is very thin. Since
scroll turbulence is essentially three-dimensional, it
bears no reflection on behaviour of spiral waves in truly
two-dimensional media. Hence the behaviour of scrolls in layers of a
given thickness may be effectively two-dimensional,
unaffected by the negative tension, or truly three-dimensional, with
full blown turbulence, or in an intermediate regime. The understanding of
possible intermediate regimes is thus vitally important for interpretation
of experimental data and for possible medical implications.

\sglfigure{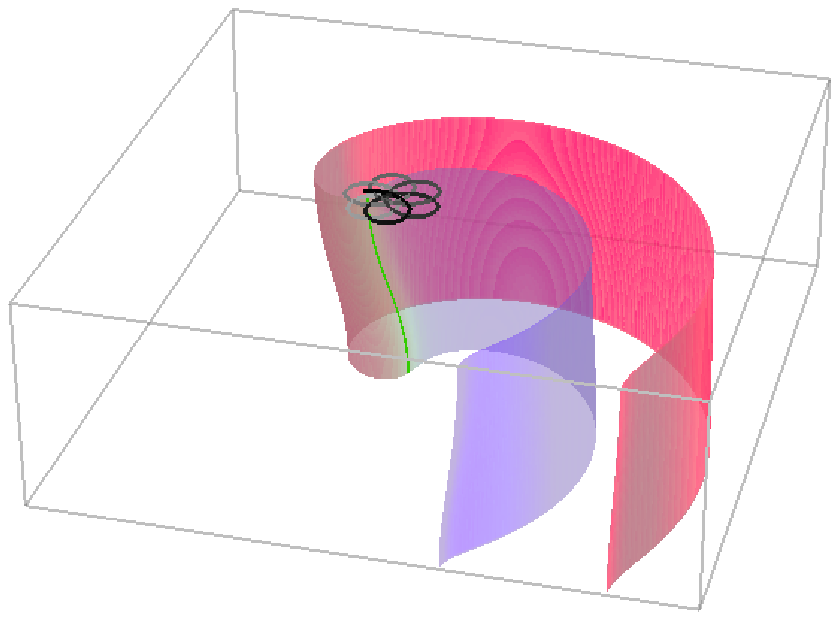}{ (color online)
  Buckled scroll and filament, with the tip path on the top of the
  box. Barkley model with $\apar=1.1$,
  $\bpar=0.19$, $\cpar=0.02$, $\Dv=0.10$, box size
  $20\times20\times6.9$~\cite{epaps}.
}{scroll}

Here we consider one such intermediate regime, which is illustrated
in~\fig{scroll}. This is a snapshot of a scroll wave solution of an 
excitable reaction-diffusion model
\begin{equation}
  \dd_t \u = \f(\u) + \D\nabla^2\u, \label{rds}
\end{equation}
where $\u,\f\in\Real^{\ell}$, $\D\in\Real^{\ell\times\ell}$,
$\u(\r,t)$ is the dynamic vector field, $\r\in\Real^3$, $\D$ is the
diffusion matrix and $\f(\u)$ are reaction kinetics that sustain
rigidly rotating
spiral waves, in a rectangular box
$\r=(x,y,z)\in[0,\Lx]\times[0,\Ly]\times[0,\Lz]$, with 
no-flux boundaries and initial
conditions in the form of a slightly perturbed straight scroll.  In
boxes with $\Lz$ below a critical height $\Lc$, the scrolls
keep straight and rotate steadily. In large enough $\Lz$, the 
scroll wave turbulence develops. In a range of $\Lz$ slightly above
$\Lc$  as in~\fig{scroll}, the straight scroll is
unstable, and its filament, after an initial transient, assumes an
S-like shape which remains constant and precesses with a constant
angular velocity.  In almost any $z=\const$ section, including the
upper and lower surfaces, one observes spiral waves with a circular
core, whose instant rotation centre, in turn, rotates with an angular
speed $\Om$, which changes little with  $\Lz$, but with a radius
which is vanishingly small for $\Lz\gtrapprox\Lc$ and grows with
$\Lz$. The resulting tip path, observed on the upper and lower
surfaces, is similar to classical two-periodic
meander~\cite{Zykov-1986}. %
  A similar phenomenon was observed in 
  a model of heart tissue~\cite{Alonso-Panfilov-2007}.

In this Letter, we investigate the instability which leads to such
buckled, precessing filaments, using linear and non-linear theory and
numerical simulations. The instability is akin to the Euler's buckling
in elasticity, where a straight beam deflects under a compressive
stress that is large enough compared to the material's
rigidity~\cite{elasticity}.

Initial  insight  can be obtained
through linearization about a straight scroll wave
  solution $\UU$ stretched along the
$z$-axis, 
as in~\cite{Henry-Hakim-2002}.
Small perturbations $\tuu$ with wave number $\kz$ 
will evolve according to $\dd_t \tuu =\HL_{\kz} \tuu$, where
\begin{equation} \label{LKZ}
  \HL_{\kz} = \D \nabla^2 -  \D \kz^2 + \om_0 \dd_\theta + \f'(\u_0).
\end{equation}
 The scroll will be stable if all the eigenvalues to
$\HL_{\kz}$ have negative real part for all allowed wave numbers
$\kz= n\ko = n \pi/\Lz$, $n\in \mathbb{Z}$. 
Analytically,
the Taylor expansion in $\kz$ for the
critical eigenvalues $\la_+$, $\la_-$, associated to translational
symmetry,
\begin{equation}
   \la_\pm(\kz) = \pm i \om_0 - (\tensc \pm i \tenps) \kz^2 - (\rigsc \pm i \rigps) \kz^4 + \OO(\kz^6) \label{lambda}
\end{equation}
relates to overlap integrals of the translational Goldstone modes and
response functions \cite{Keener-1988,
  Biktashev-etal-1994,Biktashev-Holden-1995}, see the
Appendix~\cite{epaps}. 
With the notation of
\cite{Biktasheva-etal-2009,epaps} and $ \mathbf{\hat{\pi}} = 1 -
\ket{\VV_+} \bra{\WW^+}$, we found
\begin{eqnarray}
  \tensc + i \tenps &=& \bra{\WW^+} \D \ket{\VV_+}, \label{tension}\\
  \rigsc + i \rigps &=& - \bra{\WW^+} \D (\HL-i \om_0)^{-1} \mathbf{\hat{\pi}} \D \ket{\VV_+}. \label{rigidity}
\end{eqnarray}
Thus, a filament with negative tension $\tensc$
\cite{Keener-1988, Biktashev-etal-1994, Verschelde-2007}, can
nevertheless be stabilized by higher order terms.  We call $\rigsc$
\textit{filament rigidity}; it is an analogue of the stiffness of an
elastic beam, and has the most important stabilizing effect.  If
$\rigsc>0$, then the leading-order stability condition is
\begin{equation} \label{Lcrit}
  \ko > \kc = \sqrt{-\tensc  / \rigsc}
  \quad \Leftrightarrow \quad
  \Lz < \Lc = \pi \sqrt{-\rigsc  / \tensc}.
\end{equation} 
When $\Lz$ slightly exceeds $\Lc$, a single unstable mode with spatial
dependency $\sim \cos{\pi z/\Lz}$ will grow, causing the
filament to buckle and precess at a
rate 
\begin{equation} \label{prec}
 \Omc = \tensc \left( \tensc \rigps - \tenps \rigsc \right)/\rigsc^2.
\end{equation}
The amplitude at which the buckling filament will stabilize
requires nonlinear analysis.  Our full non-linear
treatment of this phenomenon based on the time-dependent evolution
equation for the scroll filament is rather technical,
and we defer it to another publication. Here we will
consider simplified scroll dynamics, 
with the equation of motion of the scroll filament in the 
form~\cite{Dierckx-2010}
\begin{align} \label{eom}
  (\dot{\R})_\perp    = &    \left( \tensc + \tenps \dd_\s \R  \times\right) \dd_\s^2 \R 
  - \left(  \rigsc + \rigps \dd_\s \R  \times \right) (\dd_\s^4 \R) _\perp 
  \nn\\ & 
  + |\dd_\s^2 \R|^2 \left( \nonsc + \nonps \dd_\s \R  \times\right)
  \dd_\s^2 \R,
\end{align}
where $\R(\s,t)$ is filament position and $\s$ is arc length.  The coefficients $\nonsc, \nonps$ improve the phenomenological
ribbon model proposed in \cite{Echebarria-2006}; they relate to the
accelerated shrinking of collapsing scroll rings.
Linearization of Eq.~\eq{eom}  agrees with Eqs.~\eq{Lcrit} and \eq{prec} above.
A filament obeying Eq.~\eq{eom}  at $\Lz \approx \Lc$ can
be represented, in a 
in a frame precessing with
frequency $\Om$,
by its Fourier expansion $[X',Y',Z'] = [A \cos(\ko z),0,z] + \dots$
with $\ko= \pi/ \Lz$. Then collecting the terms $\sim\cos \ko z$
gives
\begin{equation} \label{ampeq}
\dot{A} = - \ko^2
A \left[ (\tensc + \rigsc \ko^2) + \ko^2 A^2 q(\ko) \right] = 0,
\end{equation}
where
$q(\ko) = -\tensc /2 + (3\nonsc/4-\rigsc) \ko^2$,
which describes a pitchfork bifurcation.
By evaluating $q(\kc)$, one finds that 
the case
$\nonsc > 2 \rigsc / 3$ yields a supercritical
 bifurcation, with stable branch 
\begin{equation} \label{amp}
A_* \approx
  \frac{\Lc}{\pi}
  \sqrt{\frac{8 \rigsc}{3 \nonsc-2 \rigsc}}
  \sqrt{\frac{L-\Lc}{\Lc}} 
  , \qquad \Lz\to\Lc.
\end{equation}
In the opposite case, the bifurcation is subcritical.

So, in absence of other instabilities (say two- or three-dimensional
meander), and subject to the inequalities $\tensc<0$, 
$\rigsc>0$ and
the limits of small $|\tensc|$ and small $|\Lz-\Lc|$, we have 
 an approximate 
solution~\eq{Lcrit}-\eq{prec}, \eq{amp}.  The
condition of negative tension, $\tensc<0$, is 
 the
key cause of the buckling instability. The condition 
$\rigsc>0$
ensures that fourth-order arclength derivatives are sufficient to
suppress high-wavenumber perturbations and so is
important only for particular formulas but not for the
phenomenon itself.
 Violation of the supercriticality condition 
$\nonsc > 2 \rigsc /3$
does not preclude the unstable branch from
becoming stable at larger $A$, as will be seen in \fig{others}(c) below.
Finally, the conditions $|\tensc|\ll 1$, $|\Lc-\Lz|\ll 1$ are 
only required for the asymptotics; in reality, one would
expect some finite, inter-dependent ranges for $\tensc$ and $\Lz$ to
support buckled scrolls. Hence we expect that buckled scrolls are
fairly typical and have ``finite chances'' to be observed in some
range of $\Lz$, if only $\tensc<0$.

In our numerical simulations~\cite{epaps} below, the asymptotics
for $\la_+(\kz)$ are evaluated using
Eqs.~\eq{tension}-\eq{rigidity},
after numerically obtaining the
modes $\ket{\VV_+}$ and $\bra{\WW^+}$ 
using \dxspiral~\cite{Biktasheva-etal-2009,dxspiral}. 
These asymptotics are compared to the numerical
 continuation of $\HL(\kz) \VV({\kz}) =
\la_+(\kz) \VV({\kz})$ 
by the parameter $\kz$ \cite{epaps}. 

We used
the reaction-diffusion system~\eq{rds} with
Barkley~\cite{Barkley-1991} kinetics, $\ell=2$, $\u=(u,v)$,
$\f=(f,g)\T$, $f=\cpar^{-1}u(1-u)(u-(v+\bpar)/\apar)$, $g=u-v$, and
$\D=\diag(1,D_v)$. We mostly use kinetic parameters $\apar$, $\bpar$,
$\cpar$ as in~\cite{Alonso-etal-2004}, which give negative filament
tension $\tensc<0$, and consider also $\Dv>0$ so as to make $|\tensc|$
smaller; note that $\Dv=1$ guarantees $\tensc=1>0$.

\sglfigure{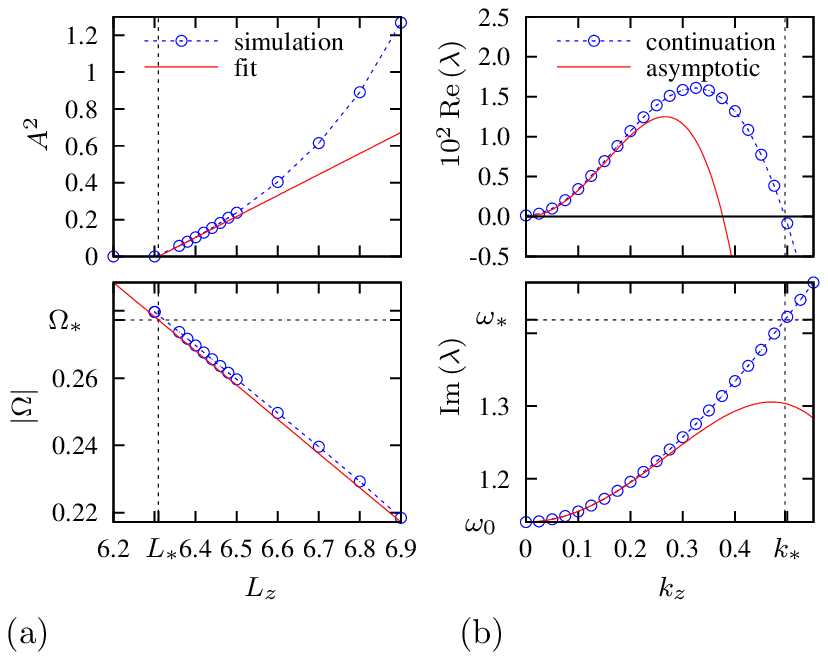}{ (color online)
  (a) Bifurcation diagram ($\apar=1.1$, $\bpar=0.19$,
  $\cpar=0.02$, $\Dv=0.1$): the amplitude (upper panel) and precessing
  frequency (lower panel) of the straight and buckled scrolls. 
  (b) The corresponding translational branch: real part (upper panel)
  and imaginary part (lower panel). For the meandering
    mode, $\Re{\lambda}<-0.24$~\cite{epaps}.
}{good}

\Fig{good}(a) shows how the buckling amplitude and precession
frequency depend on the thickness of the layer, $\Lz$, for the same
set of parameters as used to generate~\fig{scroll}. We see that
just above the bifurcation point, $\Lz\gtrapprox\Lc$, there is good
agreement with Eq.~\eq{amp}. Linear fitting of the $\Amp^2(\Lz)$
dependence for the weakest buckled scrolls gives a bifurcation point
$\Lc\approx 6.310$, and a linear extrapolation of the precessing
frequency from the same set gives $\Om(\Lc)\approx0.2789$. Panel (b)
shows the results of the linear analysis, both asymptotic as given
by Eq.~\eq{lambda} and obtained by numerical continuation of the
eigenvalue problem.
The latter gives the $\kc\approx0.497$, \ie\ $\Lc=\pi/\kc\approx6.33$,
in agreement with the direct simulations shown in panel (a). The
\dxspiral\ calculations using Eqs.~\eq{tension}-\eq{rigidity} give
$\tensc = -0.353$, $\rigsc\approx2.49$, resulting in
$\kc\approx0.376$. The nearly 25\% difference between the continuation
and asymptotic predictions is consistent with $\kz$ being not very
small, and should decrease for smaller $|\tensc|$. This is indeed
true, as seen below. The precessing frequency predicted by
continuation is $\Omc=\Im{\lamv(\kc)}-\om_0\approx
1.4188-1.1408=0.2780$, in agreement with simulations.

\sglfigure{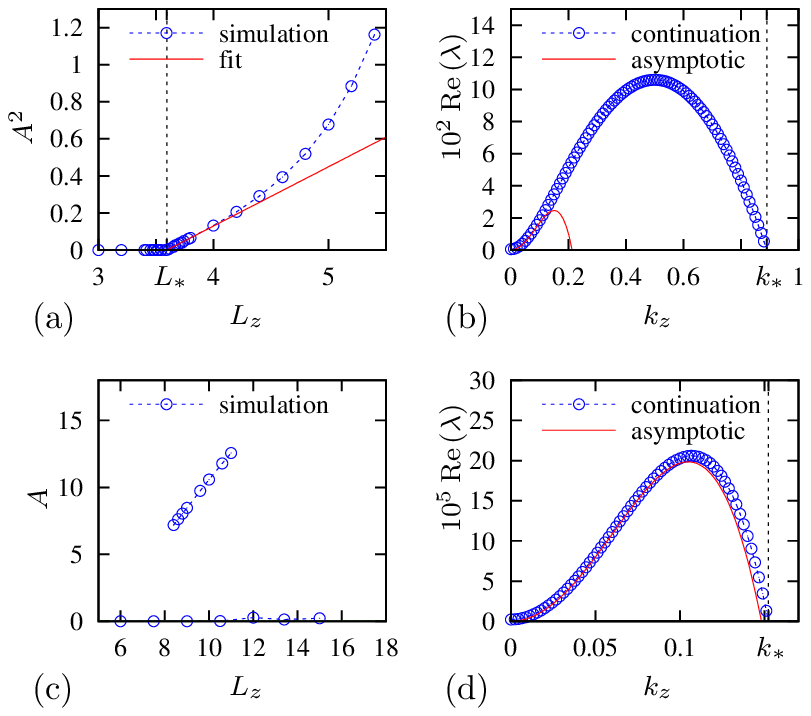}{ (color online)
  (a) Bifurcation diagram (buckling amplitude) and (b) translational
  branch (real part), for $\apar=1.1$, $\bpar=0.19$, $\cpar=0.02$,
  $\Dv=0$.  (c,d) Same, for $\apar=1.1$, $\bpar=0.19$, $\cpar=0.02$,
  $\Dv=0.25$. For the meandering
    modes, $\Re{\lambda}<-0.098$ and -0.32 respectively~\cite{epaps}.
}{others}

\Fig{others} illustrates variations in the buckling bifurcation caused
by change of parameter $\Dv$. 
 In panels (a,b),
parameters are as in~\cite{Alonso-etal-2004}
and the filament tension is strongly negative.
The \dxspiral\
predictions are $\tensc\approx-2.18$, $\rigsc\approx48.2$ so the
asymptotic $\kc\approx0.213$ is vastly different from the
continuation prediction $\kc\approx0.890$, and this discrepancy is
clearly visible in panel (b). Yet, panel (a) shows that the
bifurcation still takes place, and the critical thickness
$\Lc\approx3.60$ is in agreement with the prediction of
the continuation, $\Lc=\pi/\kc\approx3.53$. This confirms that the
assumption of smallness of the negative tension is only technical and
does not preclude buckled scroll solutions, which
still occur via a supercritical bifurcation as the medium thickness
varies. 

Panels (c,d) present a variation where the negative filament tension
is much smaller. Panel (d) shows much better agreement
between the asymptotics: $\tensc\approx-0.0362$, $\rigsc\approx1.65$,
such that $\kc=0.148$ ($\Lc = 21.23$), whereas continuation gives
$\kc=0.152$ ($\Lc=20.66$). However, the bifurcation in this case is subcritical with
a hysteresis, see panel (c), which shows that the assumption of
supercriticality is not absolute, and that a subcritical bifurcation
can likewise lead to buckled scroll solutions.

\sglfigure{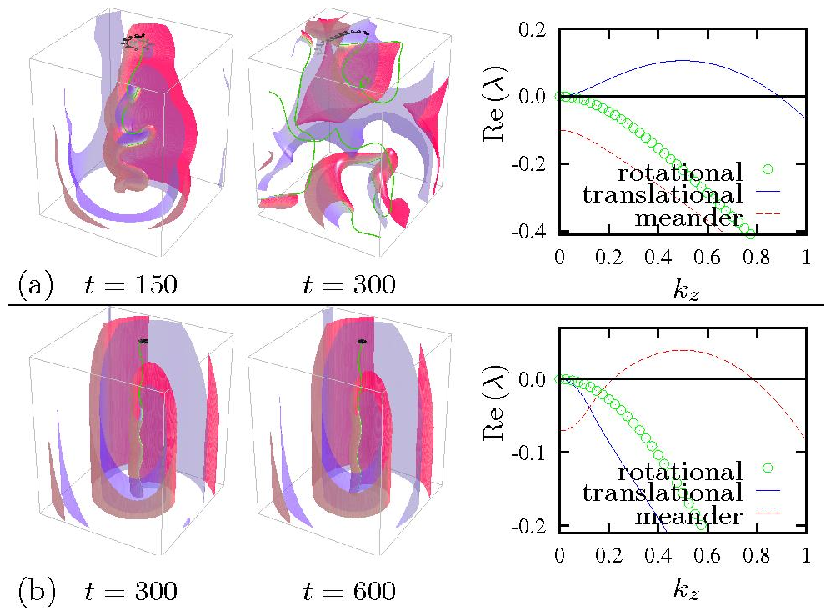}{(color online)
  Development of (a) autowave turbulence ($\apar=1.1$, $\bpar=0.19$,
  $\cpar=0.02$, $\Dv=0$) and (b) ``wrinkled scroll'' as restabilized
  solution after 3D meandering bifurcation %
    ($\apar=0.66$, $\bpar=0.01$, $\cpar=0.025$, $\Dv=0$, which
    corresponds to the leftmost point of fig.10(a)
    in~\cite{Henry-Hakim-2002}).%
   Wavefronts are cut out by clipping planes halfway through the
  volume, to reveal the filaments.  %
      Curves on the right are real parts of rotational, translational
      and meandering eigenvalue branches of $\HL_{\kz}$ from Eq.~\eq{LKZ}.
}{awme}

Finally, we illustrate the difference of the buckling bifurcation we
have described here, from the ``3D meander'' bifurcation described
previously \cite{Aranson-Mitkov-1998,Henry-Hakim-2002}. On the formal
level, the restabilized scrolls following a 3D meandering instability
look similar: at any moment, the filament has a flat sinusoidal shape
(given Neumann boundary conditions), and the top and bottom surfaces,
as well as almost every $z=\const$ cross-section,
show meandering spiral wave pictures. However, the behaviour is
completely different as $\Lz$ grows, as illustrated
in~\fig{awme}. Row (a) shows that in the negative tension case, at
sufficiently large $\Lz$ the scroll buckles so much it breaks up and a
scroll turbulence develops, in agreement with previous results. Row
(b) shows that in case of 3D meander, the amplitude remains bounded,
and even when $\Lz$ is large enough to hold several wavelengths of the
curved filament, the restabilized ``wrinkled scroll'' can persist for
a long time (compare these with ``zigzag shaped filaments'' described
in~\cite{Luengviriya-etal-2008}).  Moreover, these two bifurcations
occur in different parametric regions via different mechanisms. %
The key diffrence is, apparently,
the availability or not of infinitely small unstable
wavenumbers~\cite{epaps}.

To summarize, we predict that in an excitable medium with negative
nominal filament tension $\tensc$, a sufficiently thin
quasi-two-dimensional layer will nonetheless support transmural
filaments which are straight and stabilized by filament rigidity. When
the medium thickness $\Lz$ is increased beyond a critical thickness
$\Lc$, scroll waves may buckle and exhibit an S-shaped, precessing
filament. On the surface of the layer this will look like a classical
flower-pattern meander. If the system parameters yield a bifurcation
of the supercritical type, a stationary buckling amplitude proportional to
$\sqrt{\Lz-\Lc}$ will be reached, at which non-linear filament dynamics
compensates for the negative  tension $\tensc$. In the
subcritical case, 
loss of stability of straight scrolls will be abrupt, but it still may 
lead to restabilized buckled scrolls.
The knowledge about the buckling transition and its properties is
important for the planning and interpretation of experiments where the
medium thickness is comparable to the typical length scale of the
spiral wave. 
In particular, it can be expected that 
stability of transmural scroll waves in atrial and right ventricular
cardiac tissue may in some cases depend on filament rigidity.

\textbf{Acknowledgments} The authors are grateful to I.V.~Biktasheva
for helpful discussions. The study was supported in part by EPSRC
grant~EP/D074789/1 and EP/I029664/1 (UK). H.D. acknowledges the
FWO-Flanders for personal funding and providing computational
infrastructure.

%
\cleardoublepage
\appendix 
\onecolumngrid
\section{Appendices to ``Buckling of scroll waves '' \\ 
by H.~Dierckx,  H.~Verschelde, \"O.~Selsil and V.N.~Biktashev} 
\twocolumngrid

The first appendix is of theoretical nature; it clarifies bracket
notation and offers a proof for the rigidity coefficient expression
\eq{rigidity}. The second appendix provides details on the numerics
and processing of simulation data. The third appendix describes
  extra results that might be of interest for some readers.
  Enumeration of equations and figures here is continued from that of
  the main paper, but the list of references is separate.

\section{A. Supplementary material on theory}

\subsection{A.1 Bracket notation for Goldstone modes and response functions}

We find it convenient to adopt Dirac's bra-ket notation from
quantum mechanics and adapt them to the non-selfadjoint problems we
deal with here. Let $\Vspace$ be a suitably chosen linear space
of complex-valued $m$-component vector functions $\vv: \Real^2
\rightarrow \Complex^{m}$, the real part of which contains solutions to
our reaction-diffusion system. We also consider its dual space,
$\Wspace=\Vspace^*$, which corresponds to the space of
complex-valued linear functionals $W[\cdot]$ acting on $\vv \in
\Vspace$. The dual space $\Wspace$ consists of 
generalized functions $\ww: \Real^2 \rightarrow
\Complex^{m}$, which define those functionals
via
\begin{equation} \label{inner}
  W[\vv] =  
  \iint\limits_{\Real^2} \ww\H(x,y) \vv(x,y) \,\d{x} \,\d{y}
  =  \braket{\ww}{\vv}. 
\end{equation}
So we write functions from $\Vspace$ as ket-vectors, and functions
from $\Wspace$ as bra-vectors, assuming the scalar product between
them when bra-vector is followed by a ket-vector. 

An operator $\AA$ acting in $\Vspace$ has its adjoint operator
$\AA\adj$
acting in $\Wspace$, so that for all $\vv\in\Vspace$ and
$\ww\in\Wspace$, 
\[
  \braket{\AA\adj\ww}{\vv} = \braket{\ww}{\AA \vv},
\]
which is then briefly written as 
$\bra{\ww}\AA\ket{\vv}$.

In the context of spiral waves, one often linearizes the
reaction-diffusion system~\eq{rds}
\begin{equation}
  \df{\u}{t} = \f(\u) + \D\nabla^2\u, 		\label{rdsapp}
\end{equation}
around a rigidly rotating spiral wave solution $\UU$,
in the frame that rotates with the spiral, to find
\begin{align}
 \HL &= \D \nabla^2 + \omega_0 \dd_\theta + \f'(\UU), \nn\\
 \HL\adj &= \D\T \nabla^2 - \omega_0 \dd_\theta + \f'(\UU)\T, \nn
\end{align}
where $\omega_0$ is spiral rotation rate and $\theta$ is the polar angle.
The Euclidean symmetry of the reaction-diffusion system~\eq{rdsapp}
endows $\HL$ with three critical eigenvalues
\[
 \HL\VV_{(n)} = \la_{(n)}  \VV_{(n)}, 
 \;
 \la_{(n)} = i n \omega_0, 
 \;
 n \in \{-1,0,1\} .
\]
The critical eigenvectors $\VV_{(1)}$, $\VV_{(-1)}$ which are written
$\ket{\VV_+}$, $\ket{\VV_-}$ here, are sometimes called the
translational Goldstone modes; they can be taken in the form
\[
 \ket{\VV_{\pm}} = -\frac{1}{2}\ket{\dd_x \UU \pm i \dd_y \UU} .
\]

The spectrum of $ \HL\adj$ is the complex conjugate to the
spectrum of $\HL$, and in particular
\begin{equation}
 \bra{\HL\adj  \WW^{(n)}} = \bra{\bar\la_{(n)}  \WW^{(n)}},
 \quad \la_{(n)} = i n \omega_0
\end{equation}
(this is actually a nontrivial mathematical fact, see \eg\ the
  discussion in \cite{Biktasheva-etal-2009-A}).
A common choice of normalization is such that
\begin{equation}
   \braket{\WW^{(m)}}{\VV_{(n)}} = \delta^m_n, 
   \qquad
   m,n \in \{-1,0,1\}.
\end{equation}
The modes $\bra{\WW^+}$, $\bra{\WW^-}$ are known as `response
functions for translation'.  Their belonging to $\Wspace$ implies
  that they are effectively localized in space so that integrals like
  \eq{inner} always converge even though typical functions $\vv$ are
  only bounded but not localized. Again, see
  \cite{Biktasheva-etal-2009-A} for a more detailed discussion.

\subsection{A.2 Linearized theory}

Here we prove the result~\eq{rigidity}, which expresses the filament
rigidity coefficients $\rigsc, \rigps$ in terms of response functions.
We shall make use of a non-selfadjoint version of
the Feynman-Hellman theorem, which states how the
eigenvalue corresponding to a normalized eigenstate of
a self-adjoint operator changes upon the variation of a real-valued
parameter. 
For a non-selfadjoint operator $\AA$, if 
$\AA\ket{\VV}=\la\ket{\VV}$,
$\bra{\AA\adj\WW}=\bra{\bar\la\WW}$ and
$\braket{\WW}{\VV} = 1$
for all $T$ from a continuous interval, 
it follows that 
\begin{equation} \label{diffort}
  \braket{\dd_T\WW}{\VV}+\braket{\WW}{\dd_T\VV}=0
\end{equation}
and
$\la = \braket{\WW}{\AA\VV} = \braket{\AA\adj \WW}{\VV}$, so
\begin{eqnarray} \label{FH}
  \dd_T  \la &=&  \dd_T \bra{\WW} \AA \ket{\VV} \nn\\
  &=& \bra{\dd_T \WW} \AA \ket{\VV} 
     +\bra{\WW} \dd_T \AA \ket{\VV}
     +\bra{\WW} \AA \ket{\dd_T \VV} \nn\\
&=&  \bra{\WW} \dd_T \AA \ket{\VV}.
\end{eqnarray}
We apply this theorem to the 
operator defined in Eq.~\eq{LKZ}, i.e. 
$\AA=\HL_{\kz} = \HL - \kz^2 \D$,
and parametrize is by $T=\kz^2$. We know about the continuos 
branches of $\ket{\VV(T)}$ and $\bra{\WW(T)}$, that at $\kz=0$ they
reduce to the translational modes, 
$\ket{\VV(0)}=\ket{\VV_+}$ and $\bra{\WW(0)}=\bra{\WW_+}$.

Close to $\kz=0$, we expand 
\[
\la_+ = i \omega_0 - \ten \kz^2 - \rig \kz^4 + \OO(\kz^6),  
\]
with yet unknown complex coefficients $\ten = \tensc+ i \tenps$, $\rig
= \rigsc+ i\rigps$ (see Eq.~\eq{lambda}). 
In terms of the parameter $\kz^2 = T$, we
will be looking for $\ten = - \dd_T \la_+$ and 
$\rig = - \frac12 \dd_T^2 \la_+$,
evaluated at $T=0$. From the Feynman-Hellman
theorem~\eq{FH}
it follows
\begin{equation} \label{FH-first}
  \dd_T \la_+ =  \bra{\WW} \dd_T \HL_{\kz} \ket{\VV}  = -  \bra{\WW} \D \ket{\VV}. 
\end{equation}
Evaluated at $\kz^2 = T =0$, this recovers the well-known expression
for the filament tension coefficient, i.e. $\ten = \tensc + i \tenps =
\bra{\WW^+} \D \ket{\VV_+} $~\cite{%
  Keener-1988-A,%
  Biktashev-etal-1994-A,%
  Verschelde-2007-A%
}. Differentiation of Eq.~\eq{FH-first} gives
\begin{equation} \label{dla1}
  - \dd^2_T \la_+(0) = 
  \bra{\dd_T \WW^+} \HP \ket{\VV_+} +  \bra{\WW^+} \HP \ket{\dd_T\VV_+}
\end{equation}
The derivatives of the eigenfunctions can be evaluated by 
differentiating
$\HL_{\kz} \ket{\VV} = \la \ket{\VV}$ with respect to $T$, delivering
\[
  (\HL_{\kz} - \la_+) \ket{\dd_T \VV} = (\HP - \dd_T \la_+) \ket{\VV},
\]
and similarly for $\bra{\dd_T \WW}$. 
At $T=0$,
we have $\la_+=\om_0$, $\dd_T\la_+=\ten$. We note that
the linear equations for $\ket{\dd_T\VV}$ and $\bra{\dd_T\WW}$ are solvable,
because their right-hand sides do not have components along the
null-space of the linear operator. Namely, it is easy to see that 
$(\HP - \ten ) \ket{\VV_+}= \hpi \HP \ket{\VV_+}$
and 
$\bra{\WW^+} (\HP - \ten ) = \bra{\WW^+} \HP \hpi$,
where
\begin{equation}
  \hpi = \left( 1- \ket{\VV_+} \bra{\WW^+} \right) \label{hpi}
\end{equation}
is the projection operator which kills the components of a vector
along the null space of $\HL - i \om_0$. 
With this in mind, we get
\begin{align}
  \ket{\dd_T \VV_+} &=
    (\HL- i \omega_0)^{-1} \hpi \HP \ket{\VV_+} + C_1 \ket{\VV_+} \label{VT} \\
  \bra{\dd_T \WW^+} &=
    \bra{\WW_+} \HP \hpi (\HL- i\om_0)^{-1}  + C_2 \bra{\WW^+} \label{WT}
\end{align}
where $C_1$ and $C_2$ are arbitrary constants, which depend on the
choices of normalizations of $\ket{\VV}$ and $\bra{\WW}$ at different values of
$T$. These choices are constrained by Eq.~\eq{diffort}, which implies
\begin{multline}
 0 = (C_1+C_2)\braket{\WW^+}{\VV_+}
\\
 + \bra{\WW^+} 
 \HP(\HL- i\om_0)^{-1} \hpi 
 \ket{\VV_+}
\\
 + \bra{\WW^+}
 \hpi (\HL- i\om_0)^{-1} \HP
 \ket{\VV_+}
\\
  = C_1+C_2,                            \label{C1C2}
\end{multline}
because of the normalization
$\braket{\WW^+}{\VV_+}=1$ 
and because 
$\hpi\ket{\VV_+}=\ket{\mathbf{0}}$ 
and 
$\bra{\WW^+}\hpi=\bra{\mathbf{0}}$ 
by definition
of $\hpi$.

Substitution of Eqs.~\eq{VT} and \eq{WT} into
Eq. \eq{dla1}, with account of Eq.~\eq{C1C2}, then delivers 
\begin{eqnarray*}
  \rigsc + i\rigps &=&  -\frac{1}{2} \dd^2_T \la_+(0)\\
   &=&  \bra{\WW^+} \HP  (\HL- i \omega_0)^{-1} \hpi \HP \ket{\VV_+}, \nn
\end{eqnarray*}
which concludes our proof of Eq.~\eq{rigidity}. Note that both
rigidity coefficients vanish for a system with equal diffusion of
variables ($\HP = D_0 \mathbf{1}$), since $\hpi \ket{\VV_+} =
\ket{\mathbf{0}}$.

\section{B. Supplementary material on numerical simulations and data processing}

\subsection{B.1 Direct numerical simulations}

We used two schemes for forward evolution of the reaction-diffusion
system, an explicit and a semi-implicit.

\paragraph{Explicit scheme: } Forward Euler in time with step $\dt$,
and 7- or 19-point appoximation of the Laplacian with step $\dx$ in
cuboid domains of size $\Lx\times\Ly\times\Lz$. We used
sequential solver EZSCROLL by Barkley and Doyle~\cite{ezscroll-A}, and
our own sequential and MPI-parallel solvers. 

\paragraph{Semi-implicit:} Operator splitting between reaction and
diffusion substeps, with the diffusion substep by Brian's
three-dimensional alternating-direction
procedure~\cite{Brian-1961-A,Carnahan-etal-1969-A} which is
unconditionally stable and second-order time accurate, implemented in
our own sequential solver.

The initial conditions were in the form of a straight scroll wave with
the filament along the $z$ coordinate, slightly perturbed: slightly
twisted ($z$-dependent rotation) or slightly bended ($z$-dependent
shift).

The details specific for different simulation series are listed in
Table~\ref{tab:dns}.  The bifurcation plots in~\figs{good}(a)
and~\figref{others}(a) each were obtained through two series of
simulations: one with fixed $\dx$ and varied $\Nz=\Lz/\dx$, and the
other, to achieve a finer tuning of $\Lz$, with fixed $\Nz$ and
varying $\dx$.

\begin{table}[h]
  \caption{Discretization parameters for direct numerical
    simulations. SI: semi-implicit; E7: explicit with 7-point
    Laplacian; E19: explicit with 19-point Laplacian. 
  }
  \centering
  \newcommand{\hl}{\\\hline}
  \begin{tabular}{|l|l|l|l|l|l|} \hline 
    Figure           & Scheme & $\dt$       & $\dx$    & $\Lx=\Ly$ & $\Lz$ \hl
    \figref{scroll}  & SI     & $1/60$      & $1/10$   & 20        & 6.9 \hl
    \figref{good}(a) & E7     & $\dx^2/12$  & $1/10$   & 16        & varied \hl
    \figref{good}(a) & E7     & $\dx^2/12$  & $\Lz/64$ & $160\dx$  & varied \hl
  \figref{others}(a) & E19    & $3\dx^2/16$ & $1/5$    & 17	   & varied \hl
  \figref{others}(a) & E19    & $3\dx^2/16$ & $\Lz/19$ & $85\dx$   & varied \hl
  \figref{others}(c) & E7     & $3\dx^2/20$ & $1/5$    & 120       & varied \hl
  \figref{awme}(a,b) & E7     & $\dx^2/12$  & $1/5$    & 40        & 50     \hl
  \end{tabular}
  \label{tab:dns}
\end{table}

\subsection{B.2 Postprocessing of simulation data}

The results of simulations were visualized using a slightly modified
graphical part of EZSCROLL, based on the Marching Cubes
algorithm~\cite{ezscroll-A}. \Figs{scroll} and~\figref{awme} show
snapshots of surfaces $u(x,y,z,t)=\uc$ at selected moments of time,
semi-transparent and coloured depending on corresponding values of
$v(x,y,z,t)$: red for smaller $v$, blue for larger $v$, with a smooth
transition at around $v=\vc$. The tip line, which approximates the
instantaneous filament, was defined as the intersection of isosurfaces
$u(x,y,z,y)=\uc$ and $v(x,y,z,t)=\vc$, and is shown in green. The path
of the end of the tip line at the upper surface, \ie\ the curve
defined by $u(x,y,\Lz,t)=\uc$ and $v(x,y,\Lz,t)=\vc$, is drawn in
grayscale, with darker shade corresponding to more recent position.
We made the traditional choice for Barkley kinetics, $\uc=1/2$ and
$\vc=\apar/2-\bpar$.

The buckling amplitude and precession were defined in two
steps. Firstly, at a sufficiently frequent time sampling $(t_n)$, say at
least 30 per period, we recorded the positions of the tip line as 
$X_{m,n}=x(z_m,t_n)$, $Y_{m,n} = y(z_m,t_n)$, with $z_m = m \dx$,
$m=0,\dots,\Nz$, and at each $t_n$, approximated the tip line by
\begin{eqnarray}
  X_{m,n} &\approx& A_x(t_n) \cos(m \pi/\Nz), \nn\\
  Y_{m,n} &\approx& A_y(t_n) \cos(m \pi/\Nz), \label{cosapp}
\end{eqnarray}
using least squares.
The resulting time series for the buckling amplitude vector 
$(A_x(t),A_y(t))$
was then
averaged through periods,
\[
 \avg{\Axy} \left(\frac{T_{n+1}+T_n}{2}\right)
  =\frac{1}{T_{n+1}-T_n}\int\limits_{T_n}^{T_n+1} \Axy(t) \,d{t}, 
\]
with the time-integral implemented using the trapezoid rule. The
periods were defined via $u$ records at a selected point,
\[
  u(\xr,\yr,\zr,T_j)=\uc, \qquad \dd_t u(\xr,\yr,\zr,T_j)>0,
\]
which was typically chosen in the box corner,
$(\xr,\yr,\zr)=(0,0,0)$. These period-averaged data were then used to
define the amplitude $\Amp=|\avg\Ax+i\avg\Ay|$ and phase
$\Phase=\arg(\avg\Ay/\avg\Ax)$ of buckling. The buckling was
considered
established when the graph of $\Amp(t)$ showed saturation, subject to
residual numerical noise. The value of $\Amp(t)$ average over a
sufficiently long ``established interval'' of time was then used for
graphs in~\figs{good}(a) (top) and \figref{others}(a). The buckling
phase was made ``continuous'',
so the difference between consecutive readings of $\Phase$ does not
exceed $\pi$,
by transformation
$\Phase(t)\mapsto\Phase(t)+2\pi N_t$ with appropriately chosen
$N_t\in\Integer$. The resulting normalized dependence was approximated
in the same established interval using least squares by a linear
function of $t$, the slope of which gave the estimate of precession
frequency $\Om$, used for~\fig{good}(a) (bottom).

For \fig{others}(c), the buckling was so strong that the filament
shape was not approximated well by Eqs.~\eq{cosapp}. There we took
instead 
$\Ax(t_n)=\frac12\left(X_{\Nz,n}-X_{0,n}\right)$,
$\Ay(t_n)=\frac12\left(Y_{\Nz,n}-Y_{0,n}\right)$ as the raw data.

\subsection{B.3 Numerical evaluation of the rigidity coefficients}

Computations of spiral wave solutions $\UU$, their angular velocity
$\omega_0$, and their translational eigenmodes 
$\ket{\VV_+}$ and $\bra{\WW^+}$ were
performed by \dxspiral\ suite~\cite{dxspiral-A} based on the method
described in~\cite{Biktasheva-etal-2009-A}, which depends on three
discretization parameters: the disk radius $\rp$, radial resolution
$\Nr$ and angular resolution $\Nt$. For the eigenvalue problems, we
used straight shift-invert Arnoldi iterations without Cayley
transform, and a Krylov dimensionality of 10. The list of computed
quantities used in previous
publications~\cite{Biktasheva-etal-2009-A,Biktasheva-etal-2010-A} had
to be extended to compute $\rigsc+i\rigps$, which involved the
quasi-inversion process $\ket{\avec}\mapsto\ket{\bvec}$, where
\[
  \ket{\bvec} = \Li \hpi \ket{\avec}
\]
with $\Li$ being the inverse $(\HL - i\omega_0 )^{-1}$ in the subspace
orthogonal to $\bra{\WW_+}$. Recall that $\hpi$ is the
projection operator to that subspace, given by Eq.~\eq{hpi}.
Although the exact $\Li$ is not defined in the whole space, its
numerical implementation is defined, albeit extremely ill-posed.
(For, the more accurate is the solution of the eigenvalue
  problems for $\ket{\VV_+}$ and $\bra{\WW^+}$, the higher is
  the condition number of $\Li$).  Therefore, this computation
presented some challenge.

To achieve satisfactory results, we applied the projection operator
before and after the inverse, each several times,
\begin{equation}
  \ket{\bvec} = \hpi^k \Li \hpi^m \ket{\avec} \label{itproj}
\end{equation}
where $k$ and $m$ were integers taken as large as to ensure that
further applications of $\hpi$ did not change the results any more at
a given floating point precision (we used 8-byte
arithmetics). Obviously, the exact $\hpi$ and $\Li$ commute, so
mutliple applications of $\hpi$ would not change the result ``in the
ideal world'', and in the real computations they minimized the impact
of the round-off errors and the magnifying effect of the ill-posed
$\Li$.

Apart from straight application of the inverse $\Li$ through LU
decomposition, we also tried iterative application of the same, a
version of GMRES method and Tikhonov regularization.

The quality of the quasi-inverse was assessed by normalized residual
\[
  \left| \left| \ket{\avec} - \HL\ket{\bvec}\right| \right| /  \left| \left| \, \ket{\avec}\right| \right|,
\]
where the norm is in $l_2$.  We found that with multiple application
of $\hpi$, the simplest method gives a satisfactory quality
(normalized residual of the order of $10^{-2}$ or less) which is not
easily improved with the other, more time-consuming methods. 

\dblfigure{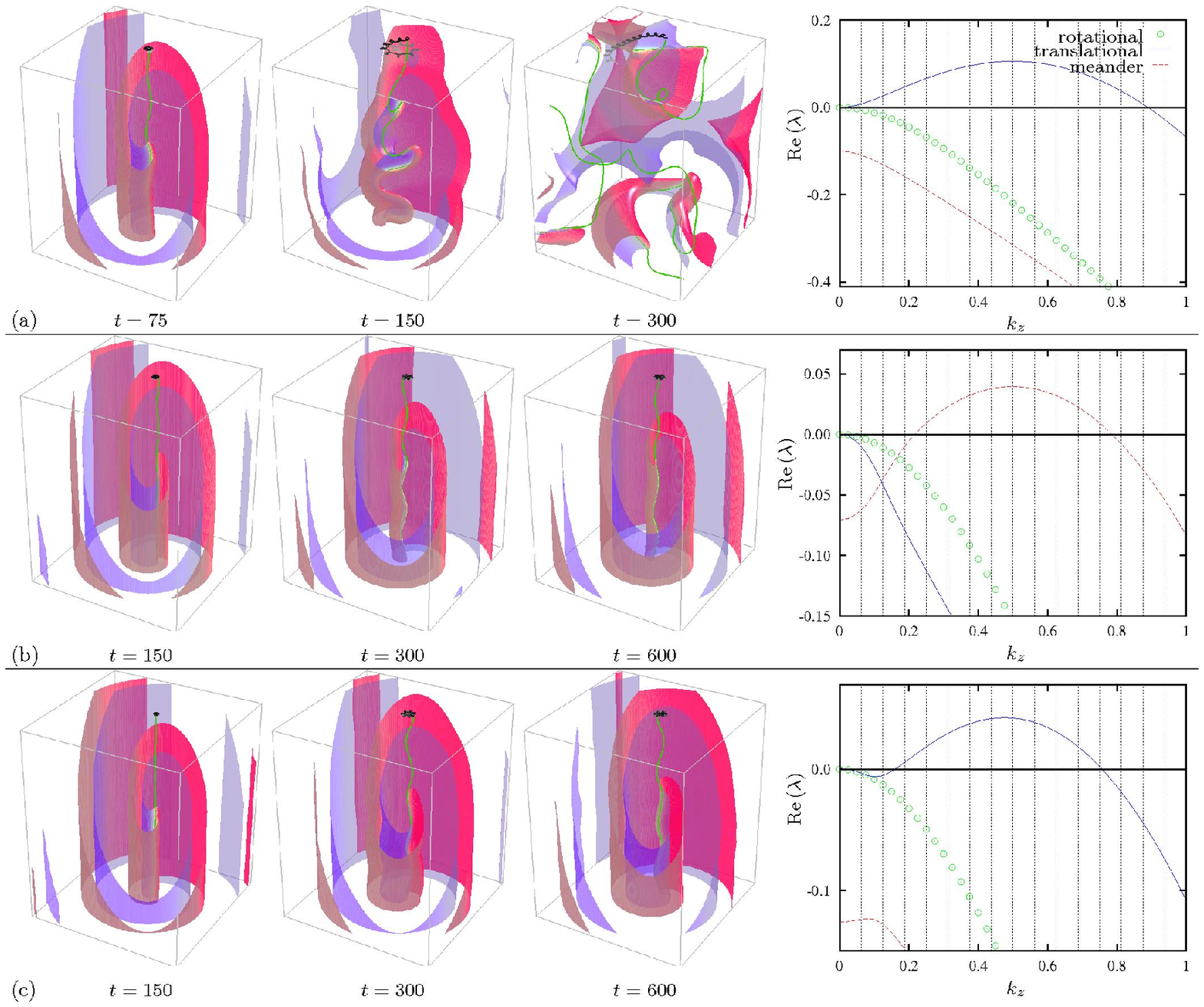}{
  Autowave turbulence vs wrinkled scrolls. 
    (a), (b) correspond to \fig{awme}; (c) $\apar=1.1$, $\bpar=0.17$,
  $\cpar=0.02$, $\Dv=0$. Box size $40\times40\times50$. The $x$-grid
  on the spectra represents the allowed wavenumbers $n\ko$,
  $n\in\Integer$, $\ko=\pi/\Lz$, corresponding to
  the given box height $\Lz=50$.
}{awwrme}

\dblfigure{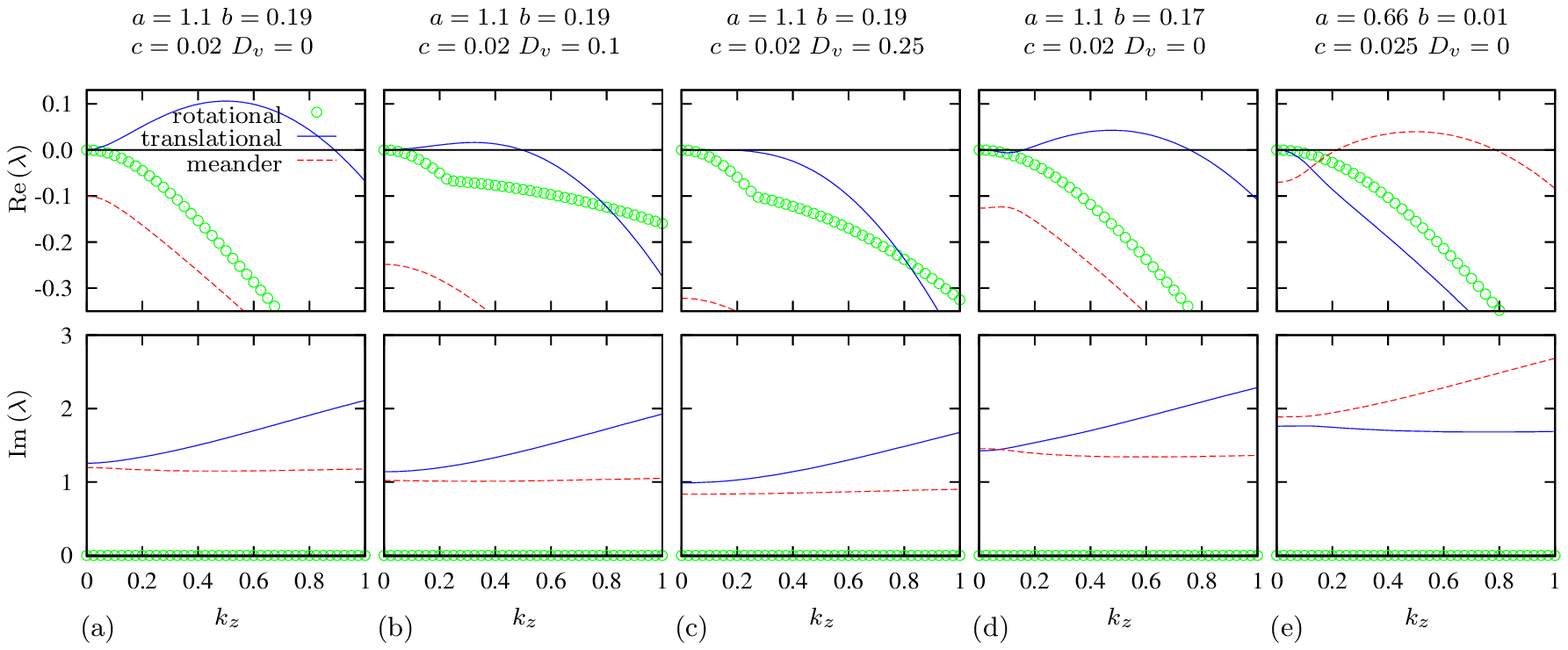}{
  Linearization spectra for all parameter sets used in the paper:
  (a) \fig{others}(a,b), \fig{awme}(a),\fig{awwrme}(a);
  (b) \fig{scroll}, \fig{good};                        
  (c) \fig{others}(c,d);
  (d) \fig{awwrme}(c);                       
  (e) \fig{awme}(b), \fig{awwrme}(b).
}{fivespectra}

\subsection{B.4 Continuation of the eigenvalue problem}

Our method is similar to that used in
  \cite{Henry-Hakim-2002-A}, up to the choice of the eigenvalue solver.
Solving the eigenvalue problem 
\[
\HL_{\kz} \ket{\VV({\kz})} = \la(\kz)\ket{ \VV({\kz})}, 
\]
by continuation in parameter $\kz$, starting
from a known  
initial value $\la(0)$,
was done at the same discretization as the unperturbed
spiral wave solution $\UU$ and the eigenmodes $\ket{\VV_+}$ and
$\bra{\WW^+}$ of the asymptotic theory. 
We used the following discretization parameters in calculating
  the eigenvalue branches: $\rp=25$, $\Nt=1000$, $\Nr=64$. The problem
  is fully resolved at these parameters, in the sense that further
  increase of either of them does not visibly change the graphs.

The rotational branch $\la_0(\kz)$ was obtained by continuation of the
  eigenvalue $\la_0(0)=0$. The translational branch $\la_+(\kz)$ was continued
  from $\la_+(0)=i\omega_0$ where $\omega_0$ was the angular
  velocity of the unperturbed spiral as found by \dxspiral.  Finding
  the starting point for the for the meandering branch $\la_m(\kz)$
  was more complicated. We have used \ezride~\cite{ezride-A} to obtain
  a steady spiral wave solution starting from cross-field initial
  conditions. The ``quotient data'', representing relative velocity
  and angular velocity of the tip, were approaching their equilibria
  in an oscillatory manner. After manually eliminating an initial transient
  period, these data were approximated by a dependency of the form
  $\Re{A e^{\lambda t}}$, $A,\lambda\in\Complex$, using Gnuplot
  implementation of the Marquard-Levenberg algorithm. Thus found
  $\lambda$ was used as an initial guess for the \dxspiral\
  calculations at $\kz=0$ and then for continuation in $\kz$ to obtain
  the meandering branch.

\section{C. Supplementary results}

\Fig{awwrme} expands on the
    comparison of the negative tension case leading to buckled scroll
    or scroll wave turbulence on one side, and the ``3D meandering''
    instability leading to ``wrinkled'' scrolls on the other
    side. Here we have added an intermediate case (lower row), which
    shows an instability of a translational, rather than meandering,
    branch, however the instability is in an interval of $\kz$
    separated from $0$. The resulting phenomenology is the same as
    with 3D meandering: a seemingly stable wrinkled scroll is
    observed. Hence it appears that for the stability of wrinkled
    scrolls it is essential that the range of unstable wavenumbers is
    separated from $0$, rather than exactly which branch shows the
    instability.  A rigorous nonlinear analysis for the two cases (b)
    and (c) would have to take into account that there are several
    unstable wavenumbers in each case.  

\Fig{fivespectra} illustrates the
    linearization spectra for all the parameter sets considered in the
    paper, shown in the same ranges for comparison. It is evident that
    spectra (a)--(d) show an instability of the translational mode,
    and spectrum (e) shows an instability of the meandering mode, and
    this is not complicated by any ``hybridization'' described
    in~\cite{Henry-Hakim-2002-A}. The only evident hybridization is of
    the rotational mode, appearing as fracture points of the
    corresponding $\Re{\la}$ curves on panels (b) and (c). Note that
    for rotational branch $\Im{\la}\equiv0$.

%

\end{document}